\documentclass[a4paper,11pt]{article}
\pdfoutput=1 

\usepackage{jheppub} 

\usepackage[T1]{fontenc} 

\usepackage[utf8]{inputenc}
\usepackage[english]{babel}
\usepackage{indentfirst}
\usepackage{fancyhdr}
\usepackage[final]{showkeys}
\usepackage{graphicx}
\usepackage{subfig}
\usepackage{amssymb}
\usepackage{comment}
\usepackage{latexsym}
\usepackage{amsthm}
\usepackage{amsmath}
\usepackage{amsfonts}
\usepackage{mathtools}
\usepackage{esvect}
\usepackage{pgf,tikz}
\usepackage{mathrsfs}
\usetikzlibrary{arrows}
\usepackage{enumerate}
\usepackage{epigraph}
\usepackage{floatpag}
\usepackage{comment}
\citeindextrue

\newcommand{\braket}[1]{\ensuremath{\left\langle #1 \right\rangle}}

\begin{document}

\title{\boldmath Critical $1$- and $2$-point spin correlations for the $O(2)$ model in $3d$ bounded domains
}

\author[a,1]{Alessandro Galvani,}
\note{Corresponding author.}
\author[b,c]{Giacomo Gori}
\author[d,c,a]{and Andrea Trombettoni}

\affiliation[a]{SISSA and INFN, Sezione di Trieste, Via Bonomea 265, I-34136 Trieste, Italy}
\affiliation[b]{Institut f\"ur Theoretische  Physik,  Universit\"at Heidelberg, Philosophenweg 19,  D-69120  Heidelberg,  Germany}
\affiliation[c]{CNR-IOM DEMOCRITOS Simulation Center and SISSA, Via Bonomea 265, I-34136 Trieste, Italy}
\affiliation[d]{Department of Physics, University of Trieste, Strada Costiera 11, I-34151 Trieste, Italy}

\emailAdd{agalvani@sissa.it, g.gori@thphys.uni-heidelberg.de,
atrombettoni@units.it}

\abstract{
We study the critical properties of the $3d$ $O(2)$  universality class in bounded domains through Monte Carlo simulations of the clock model. We use an improved version of the latter, chosen to minimize finite-size corrections at criticality, with 8 orientations of the spins and in the presence of vacancies. The domain chosen for the simulations is the slab configuration with fixed spins at the boundaries. 
We obtain the universal critical magnetization profile and two-point correlations, which favorably compare with the predictions of the critical geometry approach based on the Yamabe equation. 
The main result is that the correlations, once the dimensionful contributions are factored out with the critical magnetization profile, are shown to depend only on the distance between the points computed using a metric found solving the fractional Yamabe equation. The quantitative comparison with the corresponding results for the Ising model at criticality is shown and discussed. Moreover, from the magnetization profiles the critical exponent $\eta$ is extracted and found to be in reasonable agreement with  up-to-date results.}

\maketitle

\section{Introduction}
%
The universality class of $3d$ models with $U(1)$ [or $O(2)$] symmetry finds broad application in different areas of physics, thanks to its ability to describe various  phenomena ranging from high-energy models and lattice gauge theories \cite{Creutz1984,Rothe2010} 
to superfluids and superconductors \cite{Fradkin2013}
, while at the same time being amenable to be studied by both analytical and numerical techniques. 

The most studied lattice counterpart to the $O(2)$ field theory is the XY spin model, which can describe the chiral  transition of lattice QCD in the presence of two staggered quark flavors \cite{Kogut2006,Springer2015}. The XY Hamiltonian also provides a model for liquid helium on a lattice \cite{Matsubara1956,Betts1973}, which allows it to describe the superfluid transition of $^4\text{He}$ across the $\lambda$ line. This phase transition enjoys extremely accurate experimental studies \cite{elio}, thanks to a weakly singular compressibility of the fluid and the possibility of performing experiments in space \cite{eliospazio}, to reduce gravitational effects that would broaden the transition. However, the experimentally determined exponent of the specific heat $\alpha$ (and therefore, the exponent $\nu$ as well)  differs from the most accurate numerical results, obtained from Monte Carlo simulations \cite{Hasenbusch2019a}, by a whopping eight standard deviations. Successive conformal bootstrap results \cite{Chester2020} confirm the Monte Carlo values, meaning that this discrepancy is still an open problem. Another interesting phenomenon displayed by the XY model is the appearance of vortices and vortex lines: in $2d$, vortices drive the celebrated  Berezinskii–Kosterlitz–Thouless transition  \cite{Kosterlitz1973,kosterlitz2017nobel}, while vortex loops appear in the $3d$ XY model \cite{Shenoy1989}, providing a topological mechanism for its order-disorder transition 
an additional tool to tackle the problem of the specific heat of helium \cite{Forrester2019} and duality in (2+1)-$d$ systems at finite temperature \cite{Cvetkovic2006}.
The $3d$ XY model is such a rich  topic that a lot has yet to be discovered: in particular, we are going to focus on the properties of the critical XY model in a slab geometry.

There are many reasons to focus one's attention to systems with boundaries \citep{Diehl1997,Domb2000}: first of all, it is necessary to compare theoretical predictions with experimental results or interpret numerical data, which come from finite systems. Techniques such as finite-size scaling can be used to extract critical exponents and other important quantities. The field is also quite rich: a given bulk universality class corresponds to several surface universality classes. In some cases, boundaries are not just a nuisance, but are rather the source of the phenomenon being considered. The most well-known case is the thermodynamic Casimir effect: it has been studied thoroughly to obtain universal scaling function and Casimir amplitudes \cite{Vasilyev2009}. The interest in the Casimir effect has shifted the attention from semi-infinite systems \cite{Bray1977,Diehl1980,Diehl1981,Lubensky1975} to the geometry of a slab \cite{Grueneberg2008,Diehl2017,Gambassi2006,Gambassi2006a,Vasilyev2007}. 

The XY model in a slab, in particular, can model the Casimir effect in Helium near the superfluid transition. Different kinds of boundary conditions can be used to describe the various surface universality classes \cite{Cardy1996}. In an ordinary transition, the bulk orders in the presence of a disordered surface: this corresponds to Dirichlet boundary conditions  and is used to describe a pure $^4$He fluid, whose wave function vanishes at the boundary \cite{Gambassi2009}. Enhancing the surface coupling up to a critical value causes the boundary and the bulk to order at the same temperature; this special transition is characterized by Neumann boundary conditions \cite{Vassilev2018}. Finally, if the boundary coupling is enhanced further, or a magnetic field is added at the boundaries, the surface will order at higher temperatures than the bulk. The transition of the bulk in the presence of an ordered surface is called extraordinary transition, and is associated with a diverging order parameter at the boundaries. In the presence of two ordered boundaries, we must then distinguish the case in which the respective spins are aligned (labeled $++$) from the case where they are antiparallel ($+-$). The latter case describes experiments with mixtures of two liquids, attracted to opposite surfaces \cite{Farahmand,Maciolek2004,Deng2009}. The former case has also been studied experimentally \cite{sperimentale}. We are going to focus on the $++$ case for the $XY$ model. It should be added that, while measuring bulk quantities in an extraordinary transition, it does not matter \textit{how} the surface order was achieved, whether it is by stronger couplings or by magnetic fields localized on the boundaries \cite{Domb2000}. In our case, surface order can only be obtained through surface (magnetic) fields, since the surfaces are $2d$ and the XY model has a continuous symmetry: stronger couplings alone could not produce nonvanishing magnetization. We refer to \cite{Metlitski2020} for a recent detailed analysis of the boundary phase diagram of the $3d$ $O(n)$ model for varying $n$.

Our aim is both to obtain magnetization profiles and data for two-point correlations, and to apply the theory introduced in \cite{gori} to the XY universality class, to confirm the validity of the fractional Yamabe approach. To this end, we will describe the model that was used for numerical simulation, followed by a summary of \cite{gori}. Then we will see how the theoretical predictions provided by the fractional Yamabe equation match our numerical results.


\section{$(N+1)$-state clock model}\label{modelsection}

The spins in the XY model take value in $[0,2\pi)$. Continuous variables are, however, inconvenient for numerical simulations, as generating them and computing Monte Carlo weights takes longer. Universality comes to our aid, providing a way to bypass this difficulty: we can obtain the same critical exponents and correlation functions while considering a different model in the same universality class as the XY model.  A good candidate is the $N$-state clock model: it has the same Hamiltonian as the XY model, but the spins can take only $N$ possible values, the vertices of a regular $N$-sided polygon:
\begin{equation}
\beta H  =-\beta \sum_{<ij>}\vec{s}_i \cdot \vec{s}_j, 
\quad s_i = \left(\cos \frac{2\pi m}N,\sin \frac{2\pi m}N \right) ,\quad m\in \{ 0,\ldots,N-1 \} .
\end{equation}
\\ {It has been shown \cite{Hove2003} that the $3d$ clock model can be seen as an XY model with an additional crystal field, which reduces the symmetry of the system from $O(2)$ to $Z_N$: at the critical point, this field becomes irrelevant as long as $N>4$, meaning that it does not alter the universal properties we are seeking. In general, however, models with $Z_N$ symmetry may have different properties than models with $O(2)$ symmetry, as seen in \cite{Creutz1979}.}

In any numerical simulation, the free energy per site and other quantities contain a leading term which is the thermodynamic limit, plus various finite-size corrections.  Following \cite{Hasenbusch2019a}, we take measures to reduce these corrections: the Hamiltonian is further modified by introducing vacancies. Placing a spin in any given site lowers the energy of the system by some amount $D$; by tuning this parameter, one can find a point $D^*$ where the leading corrections to scaling vanish. The final Hamiltonian is then

\begin{equation}
\beta H=-\beta\sum_{<ij>}\vec{s}_i \cdot \vec{s}_j-D\sum_i s_i^2 , \quad \vec{s}_i \in\left\{(0,0)\right\} \cup \left\{\left(\cos \frac{2\pi m}N,\sin \frac{2\pi m}N \right) \right\}_{m=0,\ldots,N-1}.
\end{equation}

To fully define the model, it is also necessary to specify the probability measure of each spin: the $(0,0)$ state 
is chosen to have the same weight as all the other states combined, so

\begin{equation}\label{pesi}
w(\vec{s}_i)=\frac 1 2 \delta_{s_i^2,0}+\frac 1 {2N} \delta_{s_i^2,1} \ ;
\end{equation}
the partition function is then
\begin{equation}
Z=\sum_{\{\vec{s}\}}\left(\prod_{i}w(\vec{s_i})\right)e^{-\beta H}.
\end{equation}

\subsection{{Monte Carlo simulation}}

The geometry of the system is a slab of sizes $(L+1)\times 6L \times 6L$ (in the $x$, $y$, and $z$ directions respectively), with periodic boundary conditions in the parallel ($y$ and $z$) directions and fixed boundary conditions in the transverse direction. The values used for $L$ are $32,48,64,96,128$. 
The parallel directions have been chosen to be six times larger than the transverse one as they provide an excellent representation for
the system with infinite transverse directions for the observables under
scrutiny in this work. The model was simulated at the bulk critical temperature, using the values of $\beta_c=0.56379622(10)$ and $D=1.02$ obtained in \cite{Hasenbusch2019a}. We also used the same three kinds of Monte Carlo moves: two types of single-spin flip moves, and a cluster update.

The quickest spin flip move consists in proposing to empty the chosen site if it contains a spin, or fill it if empty, with a spin uniformly chosen out of the $N$ values. This move, however, does not guarantee ergodicity, which is why we also use the second, more standard, type of move. In this second case, once a site is picked, regardless of the value of the spin we generate a new spin according to (\ref{pesi}): en empty site half the time, a random spin, uniformly chosen out of the $N$ possible ones, the other half. In either case, the move is then accepted or refused with the usual criterion of the Metropolis algorithm.

One of the well-known problems of simulating systems at the critical point is the critical slowing down effect: since the autocorrelation time diverges, even after many single-spin flips we still have samples correlated to the previous ones, and therefore unfit for new measurements. This problem can be ameliorated through moves that update large (correlated) chunks of the lattice at once, using the Wolff algorithm \cite{Wolff1989}. First, we select a reflection axis by picking a vector $\vec{r}=\left(\cos \frac{2\pi m}N,\sin \frac{2\pi m}N \right)$ uniformly in $m\in \{0,\ldots,N-1\}$. Then, we pick a random spin, mark it as part of the cluster and flip it: $\vec{s}_i \rightarrow \vec{s}_i-2 (\vec{s}_i\cdot \vec{r})\vec{r}$. Next, we try and add to the cluster every neighbor of $s_i$, according to the bond probability
\begin{equation}
P(\vec{s}_i,\vec{s}_j)=1-\exp\big(\text{min}(0,2\beta \, \vec{r}\cdot \vec{s}_i \,\, \vec{r}\cdot \vec{s_j} )\big).
\end{equation}
The sites that have been added to the cluster are flipped, and the procedure continues in the same way. Once a spin has been added to the cluster and flipped, it cannot be selected or flipped again. To save time, the bond probabilities are computed once, at the beginning of the simulation.

A difference worth highlighting, with respect to the case with periodic boundary conditions, is that the cluster may reach a spin on the boundary.
To overcome this difficulty we use the recent extension of the
Wolff algorithm to the inclusion of arbitrary fields recently proposed 
in \cite{Kent-Dobias2018}. When the cluster touches the
boundary all spins on both boundary planes must be treated as a single spin $\vec{S}_B$, which gets flipped according to the same rule. All boundary spins are added to the cluster, which can then keep growing from each boundary spin.
The value of the new boundary spin is then stored to be used in future measurements. For instance, whenever the magnetization of a $yz$ plane, at a given distance $x$ from a boundary, is computed, this must be taken as the angle between the spins on the plane and the current boundary spin

\begin{equation}
m_x=\sum_{y,z}\vec{s}_{x,y,z}\cdot \vec{S}_B .
\end{equation}

The data from the simulation are used to test the hypothesis introduced in \cite{gori}, {which is summarized below.}

\section{Fractional Yamabe equation}

{The main idea is that a bounded system at the critical point alters its original (euclidean) metric in order to be as uniform as possible. The first requirement for this new metric is that the distance between any point in the system and the boundary 
be infinite, so that parts of the system that were close to the boundary while using a flat metric} now appear to be on the same footing as  ones lying deeper in the bulk. 
Next, we need to preserve local properties: the system must then locally appear euclidean, so we only allow conformal changes of the flat metric:
\begin{equation}
g_{ij}=\frac{\delta_{ij}}{\gamma(\mathbf{x})^2}.
\label{metrica}
\end{equation}
{$\gamma(\mathbf{x})$ has the role of a point-dependent length scale.
The point $\mathbf{x}=\{x_1,\ldots,x_d\}$ belongs to a $d$-dimensional bounded domain $\Omega\in\mathbb{R}^d$}. {Distances between a point $\mathbf{x}$ and $\mathbf{x}'$ measured with the metric $g$ will be written as $\mathfrak{D}_g(\mathbf{x},\mathbf{x}')$}. Requiring homogeneity in the system is implemented by making the Ricci scalar curvature constant {(the reader is reminded that $\Gamma_{jk}^i=\frac12 g^{il} \left( \partial_{k} g_{lj} + \partial_{j} g_{lk} -
 \partial_{l} g_{jk} \right)$, from which one gets the Ricci tensor
$\mathrm{Ric}_{ij} =
\partial_{l}{\Gamma^l_{ji}} - \partial_{j}\Gamma^l_{li}
+ \Gamma^l_{l\lambda} \Gamma^\lambda_{ji}
- \Gamma^l_{j\lambda}\Gamma^\lambda_{li}$ where  indices run over $\{1,\ldots,d\}$ and repeated indices are summed over)}:
\begin{equation}
    R=\mathrm{Ric}_{ij}g^{ij}=\kappa
\end{equation}
{Choosing $\kappa>0$ would mean having a space with no boundaries, such as a 
$d$-dimensional sphere $\mathbb{S}^d$, so $\kappa$ must be negative. Without loss of generality, we can set $\kappa=-1$. 
Negative curvature spaces (in their 
prototypical realization $\mathbb{H}^d$) 
indeed possess boundaries that are infinitely apart
matching our 
desiderata for uniformization.}
This requirement turns into a constraint on $\gamma(\mathbf{x})$, called the Yamabe equation \cite{Yamabe1960}

\begin{equation}\label{yamabe}
(-\bigtriangleup)\, \gamma (\mathbf{x})^{-\frac{d-2}{2} }=-\frac{d(d-2)}{4}\gamma (\mathbf{x})^{-\frac{d+2}{2}}.
\end{equation}

{There are a few cases where the Yamabe equation has simple solutions: e.g., for a ball of radius $\rho$ in any dimension, one finds $\gamma=\frac{\rho^2-|\mathbf{x}|^2}{2\rho}$ \cite{MarGonzalez2010,gori}. Solutions for the case of a slab are found in \cite{Galvani2021}.}

\begin{figure}[h]
\centering
 \includegraphics[width=.45\textwidth]{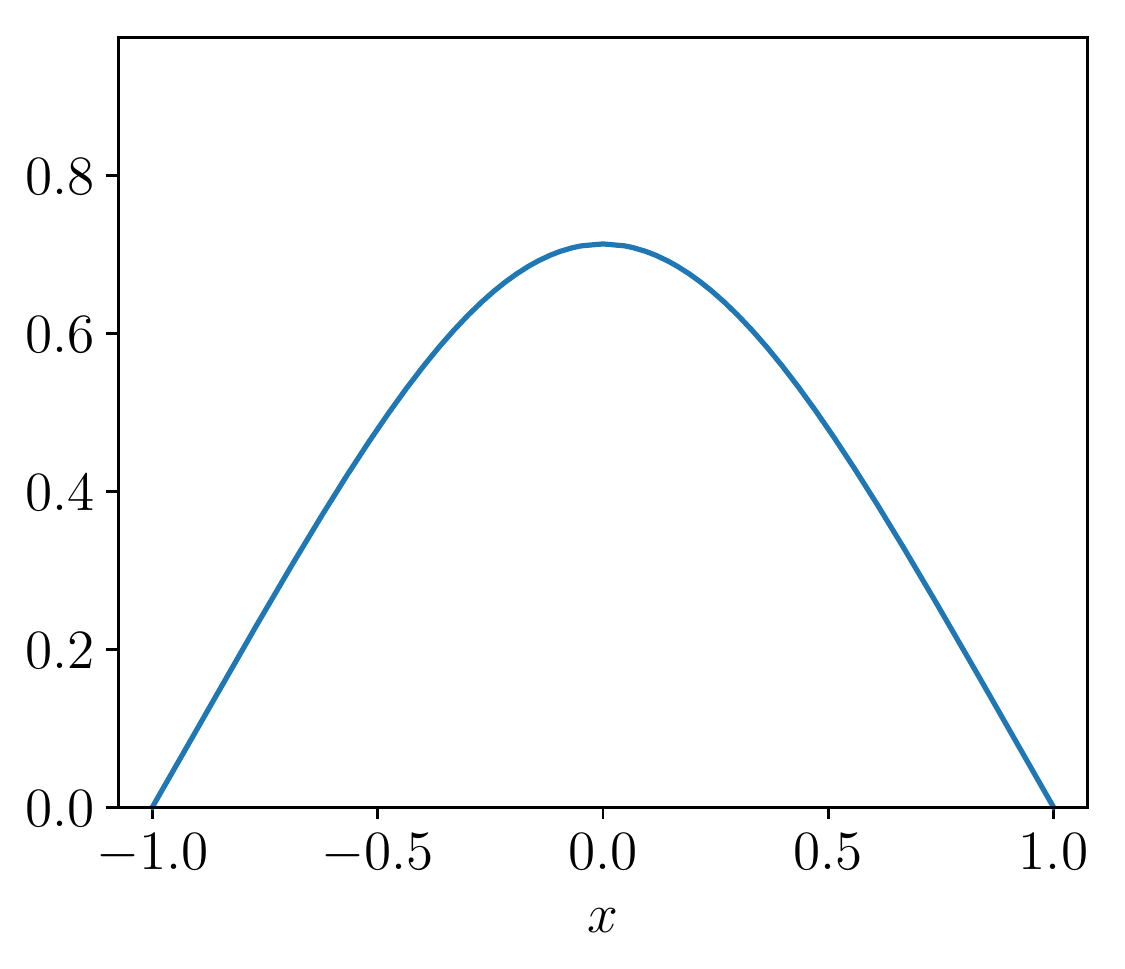}
 \includegraphics[width=.54\textwidth]{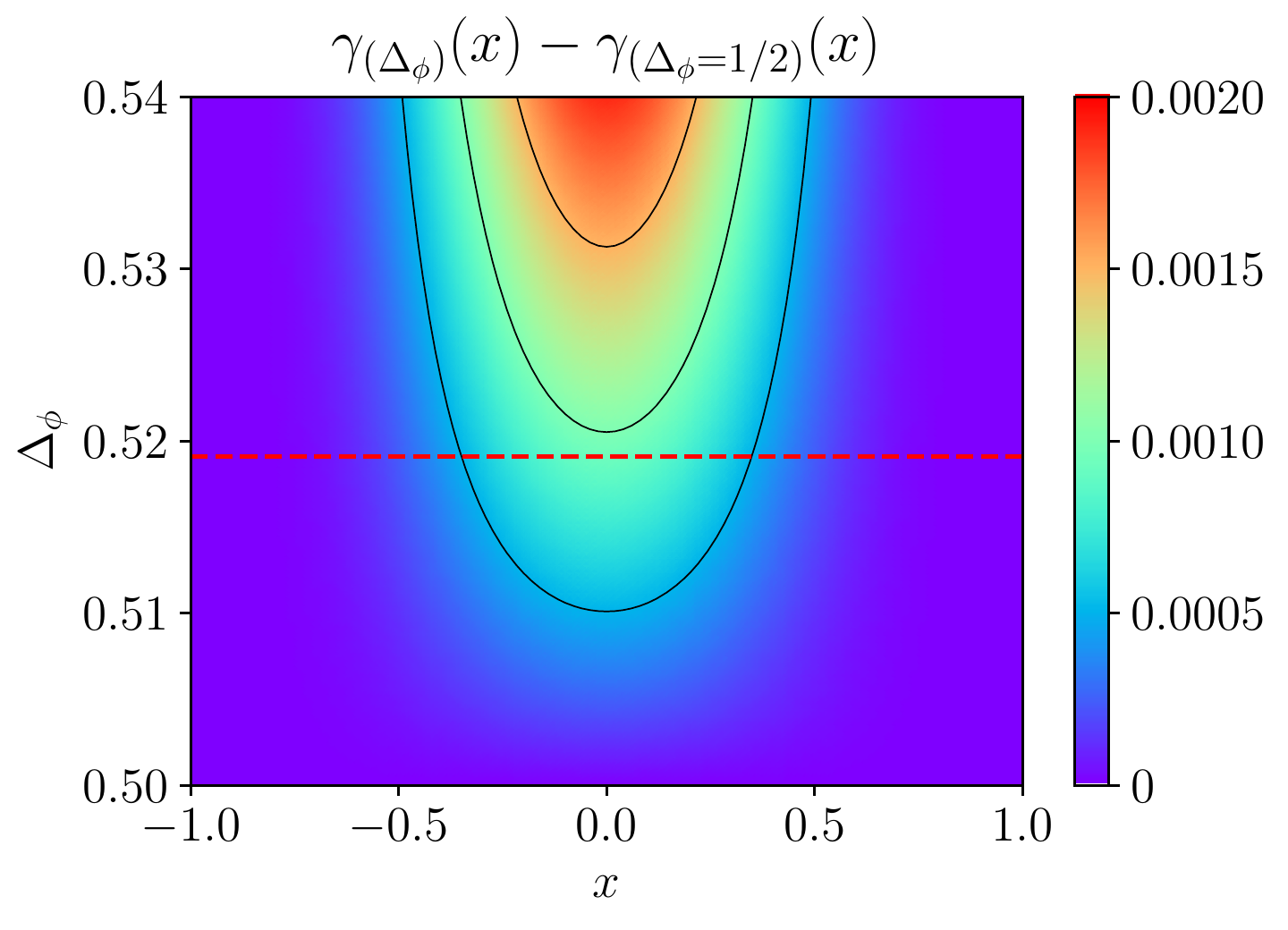}
 \caption{Solutions of the Yamabe problem in the three-dimensional slab domain ($-1<x=x_1<1$). Left: integer Yamabe. Right: solutions of the Fractional Yamabe Equation 
 as $\Delta_\phi$ is varied in the range $[0.5,0.54]$. The plot actually
 shows deviations from the $\Delta_\phi=1/2$ integer Yamabe solution (in the left pane). The red dashed line is the Conformal Bootstrap value $\Delta_\phi^{CB}$ for the XY universality class~\cite{Chester2020}.}\label{IYEandFYE}
\end{figure}

This equation, however, is only expected to be valid at the mean field level, since it does not take into account the anomalous dimension of the fields. Indeed connections between the mean-field case and the Yamabe equation have been explored and successfully tested via numerical simulations in~\cite{Galvani2021}. To give a hint, notice that the 
the exponent $\frac{d-2}{2}$ and $\frac{d+2}{2}$ correspond to the 
scaling dimensions of the order parameter field of a free theory 
in $d$ dimensions and its conjugate field respectively: {the Yamabe equation is therefore equivalent to the saddle-point equation of an $O(n)$ model at the upper critical dimension.} 
When anomalous dimensions are present, as in our case, the Yamabe equation must be modified into the fractional Yamabe equation (the dependence on the anomalous dimension of $\gamma$ has been highlighted by introducing a subscript)

\begin{equation}\label{fracyamabe}
(-\bigtriangleup)^{d/2-\Delta_{\phi}}\gamma_{(\Delta_{\phi})} (\mathbf{x})^{-\Delta_{\phi}} \propto \gamma_{(\Delta_{\phi})} (\mathbf{x})^{-d+\Delta_{\phi}}.
\end{equation}
{The introduced modification to Yamabe equation is 
in principle easy and stems from the assumption
that $\gamma_{(\Delta_{\phi})}(\mathbf{x})$ constitutes 
a local gauge for measuring lengths.}
The subtleties involved in the definition of the suitable
fractional power of the laplacian $(-\bigtriangleup)^{s}$
call upon the development of an ambient construction 
adapted for bounded domains as done in Appendix~B of~\cite{gori}. 
The fractional Yamabe equation~\eqref{fracyamabe}
can also be given some geometric meaning: it
can be interpreted as the problem of finding
a metric making $R^{(s)}$, the so-called fractional Q-curvature (of order $s=\frac{d}{2}-\Delta_{\phi}$), constant.
In this framework the ordinary scalar curvature $R$ is proportional
to $R^{(1)}$, {so for clarity one may refer to \eqref{yamabe} as the \textit{integer} Yamabe equation}. To get some intuition on why they 
could appear in a critical anomalous system consider 
that a sphere of radius $\rho$ has $R^{(s)}\propto\rho^{-2s}$.
This makes $R^{(s)}$ at least a reasonable candidate to rule 
the space dependence of a quantity with arbitrary 
real scaling dimension.
For more details on the fractional Yamabe problem
please consult~\cite{MarGonzalez2013}.
{In Figure~\ref{IYEandFYE} we collect solutions in the slab geometry of
the integer Yamabe problem and of the fractional Yamabe for a range of anomalous dimensions relevant for this work as derived in~\cite{gori}.}

Once the Yamabe equation is solved and the function $\gamma(x)$ is found, it can be used to express correlation functions~\cite{gori}: {to this end, one notices that one-point functions and the scale factor transform similarly under a dilation of the system $\Omega \rightarrow b\, \Omega$:} 

\begin{equation}
\braket{\phi_{b \,\Omega}(b\,\mathbf{x})}=b^{-\Delta_{\phi}}\braket{\phi_{\,\Omega}(\mathbf{x})}, \qquad \gamma_{b\,\Omega}(b\, \mathbf{x})=b\, \gamma_{\,\Omega}(\mathbf{x}),
\end{equation}

Therefore, once $\gamma(\mathbf{x})$ is known, all one-point functions are determined up to a multiplicative constant $\alpha$:
\begin{equation}\label{1p}
\braket{\phi(\mathbf{x})}=\frac{\alpha}{\gamma(\mathbf{x})^{\Delta_{\phi} } }.
\end{equation}

As an example, for a half-space in any dimension, with $x_1>0$ and $\{x_2, \ldots, x_d\} \in \mathbb{R}^{d-1}$,  the solution to the integer Yamabe equation is simply $\gamma(\mathbf{x})=x_1$, so
\begin{equation}
\braket{\phi(\mathbf{x})}=\frac{\alpha}{x_1^{\Delta_{\phi}}}.
\label{semisp}
\end{equation}
The hyperbolic half-space is peculiar in the sense that the same $\gamma(\mathbf{x})$ also solves the fractional Yamabe equation for any $\Delta_{\phi}$, meaning that \eqref{semisp} is also valid below the upper critical dimension.

{Higher-order correlations are not completely determined; however, this approach predicts that they will contain a dimensionful prefactor for every field, and will also depend on the distances between the points. Based on our conjecture, the bounded critical system no longer exhibits a euclidean metric: for this reasons, distances between points are to be computed with the metric $g$ we determined \eqref{metrica}: }

\begin{equation}\label{2p}
\braket{\phi(\mathbf{x})\phi(\mathbf{x}')}=\gamma(\mathbf{x})^{-\Delta_{\phi}}\gamma(\mathbf{x}')^{-\Delta_{\phi}} \mathcal{F}\left(\mathfrak{D}_g(\mathbf{x},\mathbf{x}') \right).
\end{equation}

{Both \eqref{semisp} and, for the case of the $2d$ Ising model, \eqref{2p}, reproduce known results from Conformal Field Theory, as shown in section~\ref{Isingcomparison}.} In higher dimensions, they also reproduce what 
is known for specific domains that are interiors of balls, possibly of infinite radius, reducing to semi-infinite systems (see \cite{Cosme2015,Gori2015} for applications in the spirit of this work).
For any other domain the outlined approach provides new, 
testable predictions that have already
been numerically verified in the Ising model~\cite{gori}.
The hypotheses for one-point and two-point functions will now be tested against the result of the Monte Carlo simulations for the system
under scrutiny belonging to the XY universality class.

\section{Results}

\subsection{Magnetization profiles}
\begin{figure}[h]
\centering \includegraphics[width=.8\textwidth]{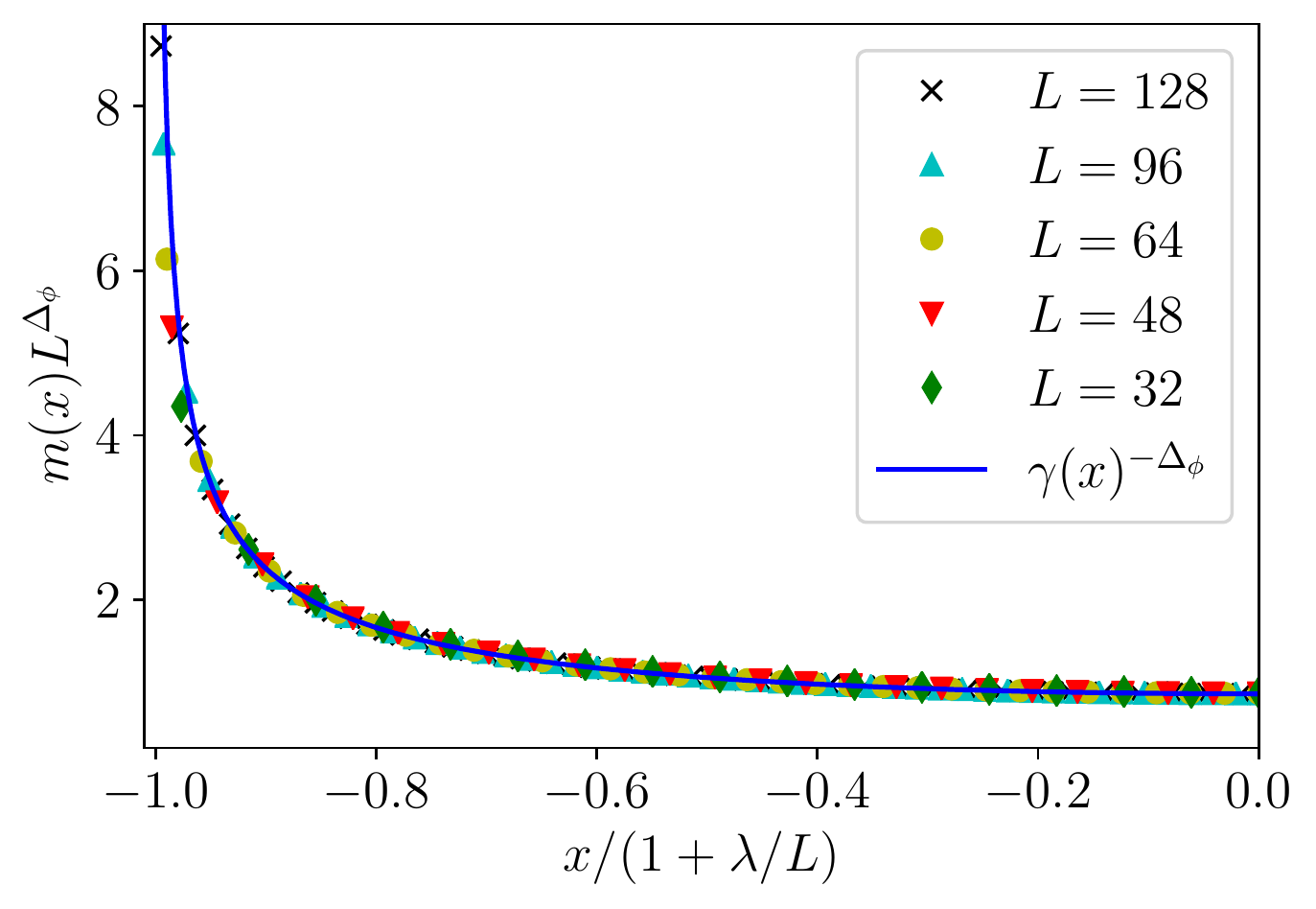}
\caption{Magnetization profiles for different sizes, rescaled by $L^{\Delta_{\phi}}$ and with the extrapolation length to show the collapse. The transverse direction of the slab is $x\in [-1,1]$; the values of the magnetization at $x$ and $-x$ have been averaged, so we plot only the left half of the profile.}\label{profilo}
\end{figure}

{To test \eqref{1p}, we employ the clock model in a $d=3$ slab defined in Section~\ref{modelsection}. We measure the average of the order parameter field, i.e., the magnetization, on various planes parallel to the boundaries, obtaining a function of the transverse distance. 
In order to improve readability in this section and in the following 
we slightly change the notation since we will be dealing just with 
models in three (two) dimensions: the coordinates of a point 
$\mathbf{p}$ will be denoted by $\{x,y,z\}$ ($\{x,y\})$. 
If there is a preferred direction it will be
chosen to be the $x$ coordinate.
The magnetization profile will thus be $m(x)$ with $x$ transverse
to the slab.
This profile can then be fitted by}

\begin{equation}\label{fiteq}
\braket{\phi(x)}=m(x)=\alpha \left[ L\,\, \gamma_{(\Delta_{\phi})}\left( \frac x {1+\lambda/L}\right)\right]^{-\Delta_{\phi}}.
\end{equation}
The fit parameters are a multiplicative constant $\alpha$, the scaling dimension of the field $\Delta_{\phi}$ and the extrapolation length $\lambda$. This last parameter is introduced to match numerical profiles for spin systems with continuous profiles from field theory: on the lattice, the magnetization does not diverge. A divergence would appear if the lattice profile were continued $\lambda$ sites beyond either boundary \cite{Cardy1996,Diehl1997}. We notice that the extrapolation length remains roughly constant as the system size is increased, meaning that larger sizes capture a greater fraction of the continuum profile. 

In Figure \ref{profilo}, we plot the magnetization profiles for different system sizes, appropriately rescaled by multiplying each profile by $L^{\Delta_{\phi}}$. The $x$ coordinate has also been rescaled, to take into account the extrapolation length of each size. 
First of all, we notice a clear collapse of the different profiles onto the same curve, proving that we are at the critical point. For each system size, the fit using \eqref{fiteq} gives us a value of the scaling dimension $\Delta_{\phi}$, as shown in Table \ref{tabelladelta}. As our final estimate, we take the average of the different values, obtaining 
    \begin{equation}\label{ourbestdeltaphi}
        \Delta_{\phi}=0.5206(15).
    \end{equation}
This is compatible, to one standard deviation, with the most precise estimate $\Delta^{CB}_{\phi}=0.519088(22)$, obtained through conformal bootstrap \cite{Chester2020}. This result should be intended as a check of the theory, since obtaining critical exponents is far from the only use of the critical geometry approach.
Details on the data analysis are found in Appendix \ref{app}.

\begin{table}[h]
    \centering
    \begin{tabular}{|c|c|}
    \hline 
    $L$ & $\Delta_{\phi}$ \\
    \hline 
    32 & 0.5227(7) \\
    48 & 0.5207(3) \\
    64 & 0.5194(2) \\
    96 & 0.5205(2) \\
    128 & 0.5193(2) \\
    \hline 
    \end{tabular}
    \caption{Scaling dimensions obtained from the fit for each system size. As final value, we take their average: $    \Delta_{\phi}=0.5206(15)$.}
    \label{tabelladelta}
\end{table}

\subsection{Correlation functions}

 Testing (\ref{2p}) is slightly more subtle because we cannot determine the function $\mathcal{F}$. Once the correlation data are divided by the product of one-point functions, i.e.,
\begin{equation}
r(\mathbf{x},\mathbf{x}')=\frac{\braket{\phi(\mathbf{x})\phi(\mathbf{x}')}}{\braket{\phi(\mathbf{x})}\braket{\phi(\mathbf{x}')}},
\end{equation} 
they will depend only on the distance between the points computed with our metric $g$. This means that the points will collapse onto a single line if plotted as a function of this distance, while they will appear scattered if plotted as a function of any other (nonequivalent) distance. The data we collected is averaged over the parallel directions:
\begin{equation}
C(x,x',\Delta y, \Delta z)= \frac{1}{(6L)^2}\sum_{y,z=1}^{6L} \braket{\phi(x,y,z)\phi(x',y+\Delta y,z+\Delta z)}.
\end{equation}

To limit the volume of data, we measured correlations only in a $15\times 15 \times 15$ grid: $x,x'=(i+1)L/16$, $\Delta y,\Delta z=iL/16$, where $i=0,\ldots, 14$. Once we take $x\geq x'$, we consider the points $C(x,x',\Delta y,\Delta z)$ and $C(x,x',\Delta z,\Delta y)$ to be the same data point, and exclude coinciding points, we get 7672 independent correlators. The fractional Yamabe distance between the points was computed via Surface Evolver \cite{Brakke1992} using as metric $g=\delta/\gamma_{(\Delta_{\phi}^{CB})}^2$.

\begin{figure}
    \includegraphics[width=.5\textwidth]{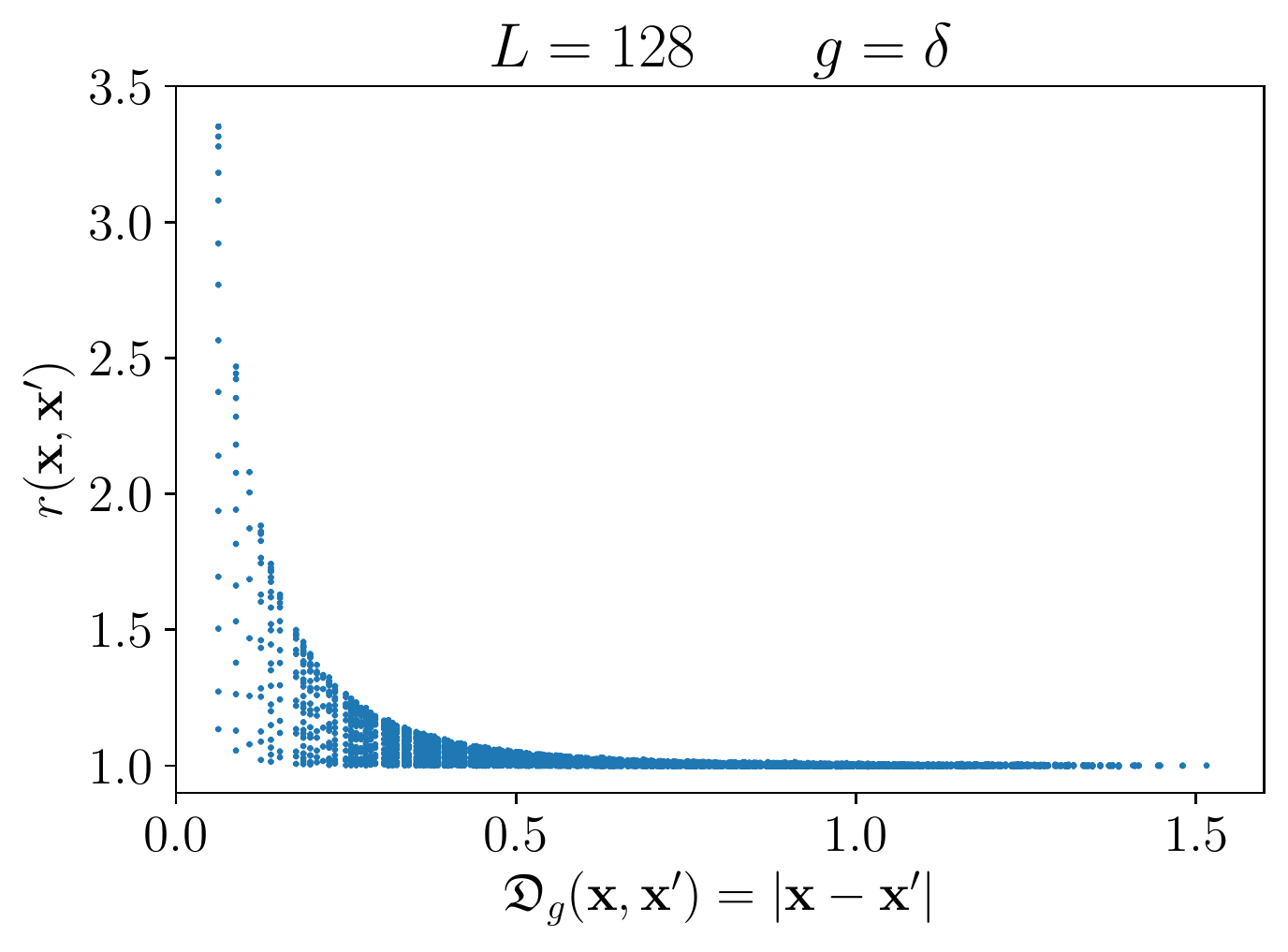}
    \includegraphics[width=.5\textwidth]{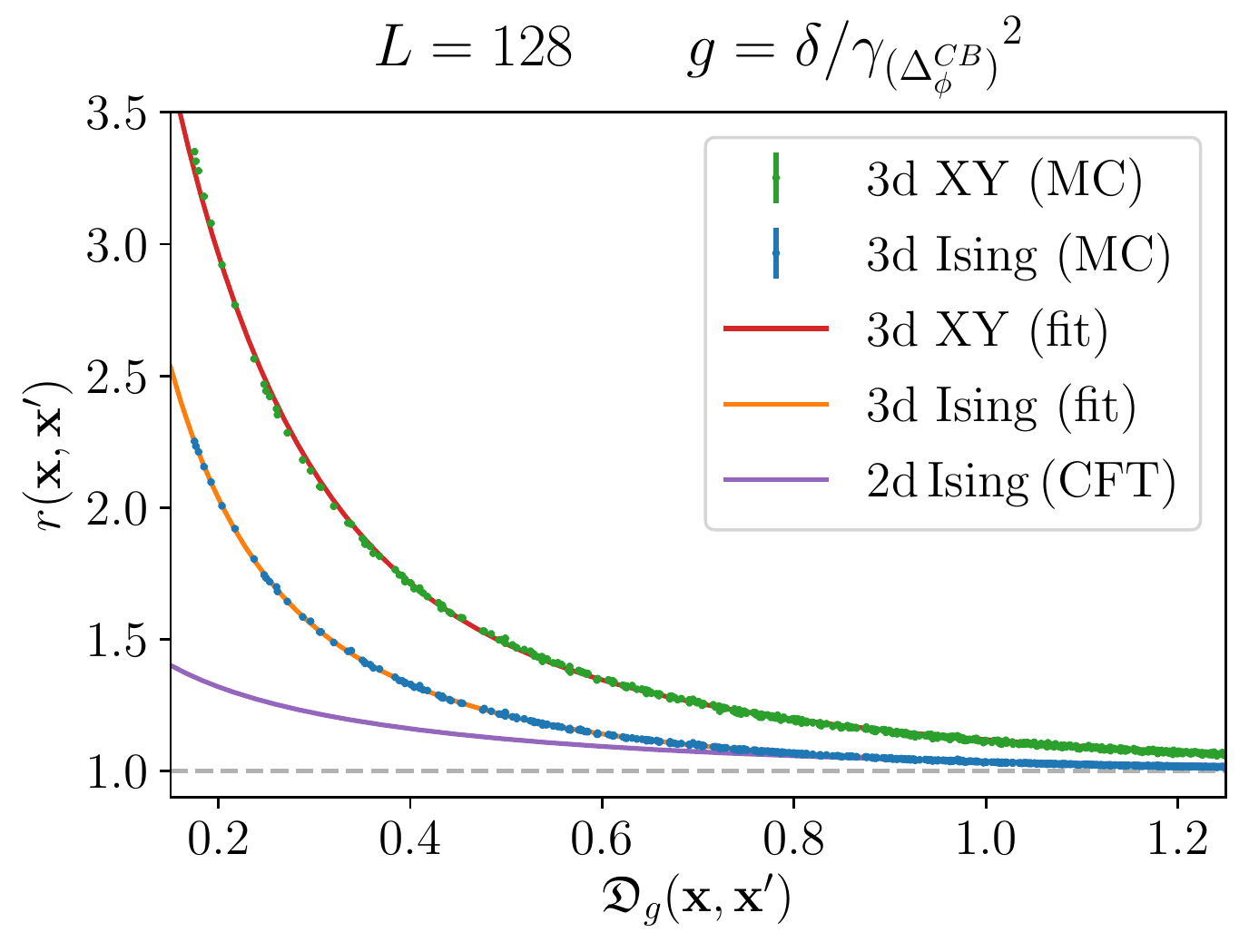}
    
    \caption{{Correlation ratio $r(\mathbf{x},\mathbf{x}')=\braket{\phi(\mathbf{x})}^{-1}\braket{\phi(\mathbf{x'})}^{-1}\braket{\phi(\mathbf{x})\phi(\mathbf{x}')}$: on the right, plotted as a function of the euclidean distance: since that is not the ``correct'' metric, the points are scattered. On the right, the same points plotted as a function of the distance $\mathfrak{D}_g(\mathbf{x},\mathbf{x}')$, computed with the metric which solves the fractional Yamabe equation for $\Delta_{\phi}=\Delta_{\phi}^{CB}$, the most accurate estimate. The points collapse onto one line. In the inset a zoom of the region where the ratio displays more pronounced variations is shown. } }
    \label{corr2}
\end{figure}

The correlation ratio as a function of the distance $\mathfrak{D}_g(x,y)$ are shown in {the left pane of} Figure \ref{corr2}: we see a clear collapse onto a single line. 
The value chosen for the scaling dimension
$\Delta_\phi$ affecting $\gamma_{(\Delta_\phi)}(x)$ and in turn the 
distances $\mathfrak{D}$ is $\Delta_\phi^{CB}$; however taking 
our best estimate~\eqref{ourbestdeltaphi} makes actually no visible difference.
To see what would happen for a different, wrong from the outset, metric, we also plotted the same points as a function of the euclidean distance between them, in {the right pane of} Figure \ref{corr2}: a collapse is absent and the point are scattered.

To quantify the goodness of the collapse, we fit the points in \ref{corr2} with the function $f(\xi)=1+a_1 e^{-b_1 \xi}+a_2 e^{-b_2 \xi}$, and then compute the mean square displacement between the points and the curve ($n_{\text{d.o.f}}$ are the independent degrees of freedom):

\begin{equation}
    \sigma=\sqrt{\frac{\left[r -f(\mathfrak{D}_g) \right]^2}{n_{\text{d.o.f}}} }.
\end{equation}
We obtain $\sigma=0.0077$ for $L=32$, $\sigma=0.0036$ for $L=64$ and $\sigma=0.0029$ for $L=128$: the fact that displacements become smaller for larger sizes 
lends quantitative support to the collapse. Notice that
the chosen fitting function $f$ correctly captures the
long (mutual) distance limit of the correlation ratio, that is 
$r\rightarrow 1$ as $\mathfrak{D}\rightarrow\infty$.
Concerning the short distance properties, the operator product expansion structure should be 
retrieved; this, however, is not the focus of the present investigation,  since short distance behavior
is not crucially affected by the curved space properties we are interested in, as well as being challenging to retrieve from lattice approaches.

\subsection{Comparison with the Ising model}\label{Isingcomparison}
As we mentioned, the function $\mathcal{F}(\mathfrak{D}_g)$ of the distance between the points is not known in $d\neq 2$ and our approach cannot determine it without using the input from the numerical simulations.
Figure~\ref{corr2} shows the existence of $\mathcal{F}$ through the collapse of the data. One sees that we can obtain 
a precise numerical estimate of it. 

\begin{figure}
    \centering
    \includegraphics[width=.9\textwidth]{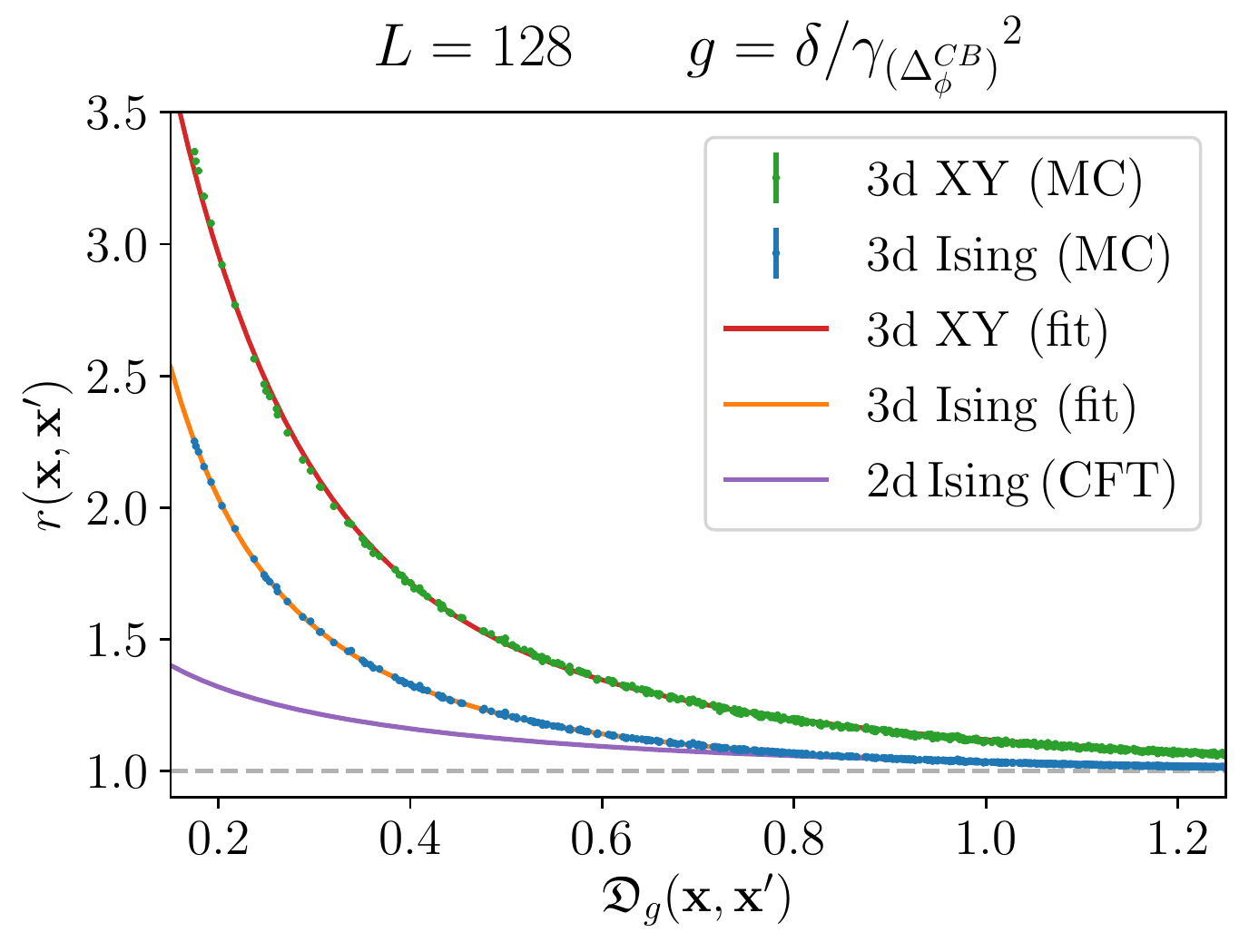}
    \caption{Correlation ratio of the $3d$ Ising and clock models, each computed as a function of the respective fractional Yamabe distance. {The lines shown under the numerical data are fitting functions. The exact function is shown also for the Ising $2d$ as obtained exactly from CFT.}}
    \label{isingxy}
\end{figure}

To put in context the result obtained for the $\mathcal{F}$ of the $3d$ XY model, we can now set it side by side with the same function for the Ising model. In Figure \ref{isingxy}, we compare the correlation ratio
$r$ we obtained for the XY class with the corresponding profiles for the $3d$ Ising model (obtained in \cite{gori}). 
We see qualitatively similar curves in the two cases, with greater correlation ratio for the XY model with respect to the Ising one. 
The quality of the collapse in $3d$ both for the XY and the Ising models is similarly convincing. 
Notice that the metric for the Ising model is different from the one for the XY model, since they have different scaling dimensions, producing distinct fractional Yamabe equations and yielding in turn different collapse distances. As expected, both correlation ratios tend to $1$ for large~$\mathfrak{D}$.

To analyze the role of the dimension $d$, we consider the $2d$ Ising case, which is an important example of the validity of our approach when applied to $d=2$. Obtaining the function $\mathcal{F}$ for the $2d$ Ising model is straightforward via boundary Conformal Field Theory (CFT). In particular, the two-point function for the Ising model on a half-plane $x>0$, with diverging magnetization at the boundary, is known to be \cite{difrancesco}:

\begin{equation}\label{halfplane}
    \braket{\phi(\{x,y\})\phi(\{x',0\})} = \frac {\alpha^2}{(x\, x')^{1/8}} \sqrt{\tau^{1/4}+\tau^{-1/4}}, \qquad \tau=\frac{y^2+(x+x')^2}{y^2+(x-x')^2},
\end{equation}
where $\alpha$ is a constant. This clearly contains the product of one point functions $\braket{\phi(\{x,y\})}=\alpha\,x^{-1/8}$, and $\tau$ is indeed a function of the hyperbolic distance between the two points, since

\begin{equation}
    \mathfrak{D}\big( \{x,y\},\{x',0\} \big)=2\,\text{arcsinh}\left(\frac 1 2 \sqrt{\frac{(x-x')^2+y^2}{x\, x'} } \right),
\end{equation}
so $\tau= \text{coth}^2 \, (\mathfrak{D}/2)$. This means that \eqref{halfplane} is in the form \eqref{2p}.\footnote{Remember that
the hyperbolic metric is $g_{ij}=\frac{\delta_{ij}}{x^2}$ and it solves the
Yamabe problem for the half-plane, and actually solves the fractional
Yamabe problem since it makes the half-plane fully homogeneous so all sensible curvatures are constant. This is discussed in detail in~\cite{gori}.} One can then conformally map the half-plane into a strip, and obtain a result which is still in the form \eqref{2p}: additionally, while one-point functions and the distances between the points change with the mapping, the function $\mathcal{F}$ remains the same.

Figure \ref{isingxy} is the main result of this work, as it shows that the critical geometry approach is valid for different models. It appears that the $2d$ case has weaker correlations (i.e., smaller $r$) than $3d$, and --- comparing models living in the same space dimension --- the XY model has stronger (normalized) correlations than the Ising one.

Simulations for the $1$-point magnetization of the $4d$ Ising model have been recently performed \cite{Galvani2021}. Preliminary results for the $2$-point spin correlations show that, at the considered sizes, collapse takes place similarly to Figure \ref{isingxy} and the correlation ratio $r$ appears to be different from $1$, although tending to $1$ for large distances, as it should. A careful analysis of the logarithmic corrections at the upper critical dimension and more numerical data are needed to draw further conclusions.

In future works, it would be interesting to compare the results in Figure \ref{isingxy} with the corresponding results for $3d$ $O(n)$ models with $n\geq 3$ and for models at the upper critical dimension. Finally, a crucial point is to confirm that, as is the case in $2d$ and as the Yamabe approach seems to suggest, the curve of Figure \ref{isingxy} in $3d$ does not depend on the bounded domain one chooses, be it a slab, a half-space, or any other.

\section{Conclusions}

We have shown that the introduction of a curved metric in a bounded domain can reliably describe correlation functions of statistical fields for the $3d$ XY universality class. Numerical simulations of a $3d$ improved clock model on a slab geometry have provided magnetization profiles consistent with the critical geometry approach  \cite{gori,Galvani2021}, as well as a clear collapse of the $2$-point functions. Our main results are summarized in Figure \ref{isingxy}, where the structure of correlators has been determined and the dependence on the curved metric distance is found for the $3d$ XY model and compared with the Ising model in $3d$ and $2d$. A value of $\Delta_{\phi}$ has been determined, and is consistent with the most recent estimates.

An open question, suggested by the results of the present paper, is whether there is a way to determine the function of the curved distance which appears in (\ref{2p}), to determine $2$-point correlations completely. One could also investigate whether correlations in  different $O(n)$ models (such as the Ising, XY and Heisenberg models) could by described by a boundary-independent function  dependent on $n$. Moreover, since in this paper we argued that 1- and 2-point spin correlation functions
in a bounded domain are described by the fractional Yamabe equation, it would be interesting to investigate possible generalizations of the Yamabe equation in connection to models with vectorial order parameters and more general symmetry groups.

The well-known $3d$ models (such as the Ising and XY models, as well as percolation) have a rather small anomalous dimension, meaning that $\Delta_{\phi}$ is close to the mean field value $1/2$, which is described by the integer Yamabe equation \cite{gori,Galvani2021}. This suggests another line of research: it may be possible to obtain an approximate solution of the fractional Yamabe equation as a perturbation of the corresponding integer equation. This would provide a perturbative solution to a fractional differential equation without the need to compute fractional derivatives.
Finally, the study of critical correlation functions and exponents through the critical geometry approach based on the fractional Yamabe equation is not limited to unitary models. A famous example of a statistical model lacking unitarity is percolation, which is therefore a  subject worth considering for future studies.

\vspace{1cm}

\textit{Acknowledgements}: The authors thank M. Hasenbusch for useful discussion and correspondence. GG is supported by the Deutsche Forschungsgemeinschaft (DFG, German Research Foundation) under Germany’s Excellence Strategy EXC~2181/1~-~390900948 (the Heidelberg STRUCTURES Excellence Cluster). GG also acknowledges QSTAR for hospitality during completion of this work. Stimulating discussions with several participants of Bootstat 2021, held in May in Institut Pascal, Université Paris-Saclay and online, are also acknowledged.

\appendix

\section{Simulation and data analysis details}
\label{app}

To ensure the validity of the simulations described in the main text, the following tests have been performed:

\begin{itemize}
    \item The critical temperature obtained in \cite{Hasenbusch2019a} has been checked by computing the Binder cumulant \cite{Binder1981} in a system without fixed boundaries.
    \item The system has also been simulated at $\beta_c+\sigma_{\beta_c}$ and at $\beta_c-\sigma_{\beta_c}$ (with $\beta_c=0.56379622$ and $\sigma_{\beta_c}=10^{-7}$): the resulting magnetization profiles are compatible with the ones obtained at $\beta_c$, i.e. they are within the respective statistical errors. This means that, for the studied sizes, the uncertainty on the critical temperature does not affect our results. However, further increasing the size $L$, the uncertainty on  the location of the critical point could affect the results, and supplementary studies would be necessary. 
    \item The profiles have been obtained for various ratios between the length in the parallel directions and in the transverse direction. The final value we chose is 6, since increasing it further does not change the profiles beyond statistical uncertainties.
    \item Additional small-size simulations were performed using only single-spin flips, without cluster update, as a cross-check: once again, we verified that the results were compatible with the ones obtained from the employed optimized algorithm, which includes cluster updates.
\end{itemize}
\vspace{.3cm}
{For any system size $L$, two-to-four-day simulations were performed on up to 160 Intel(R) Xeon(R) CPUs E5-2680 v2 cores running at 2.80GHz.}

After the magnetization data have been obtained, an additional step is needed before performing the fit. The points closest to the boundary are most affected by finite-size effects. Therefore, despite having smaller errors than the central points, a few of them have to be discarded. In order to determine how many to discard in an unbiased way, as well as to avoid a sharp distinction between discarded and included points, we introduce a window function $w(x)$. The weight of each point in the fit is given by the square of the ratio between this function and the error of that point. This function starts off from 0 at the boundary, ramps linearly to $1$ around a movable point $t$, and maintains the value $1$ up to the center of the slab. To determine the location of the point $t$, we start from $t=-1$ (the boundary point) and gradually move towards $t=0$. For each value of $t$ we compute the $\chi^2$ of our data, and the corresponding $p$-value. We stop once the $p$-value reaches the reference value of $p=0.95$. 

Once the values $\Delta_{\phi}(L)$ have been obtained, as seen in Table \ref{tabelladelta}, the final value is calculated as their average; the error on $\Delta_{\phi}$ is the standard deviation $\sigma$ of the set $\{\Delta_{\phi}(L) \}$. Since their errors are too small for them to be compatible with one another, $\sigma$ was not divided by the square root of the number of data points.

\bibliography{main}

\end{document}